%% file: main.tex
\documentclass{format/aa}

\usepackage[T1]{fontenc}
\usepackage{txfonts}
\usepackage{amsmath,graphicx,booktabs,array}
\usepackage{rotating}
\usepackage{refcount}
\usepackage{xspace}
\newcommand{\snr}{\ensuremath{\mathrm{SNR}}\xspace}
\makeatletter
\newcommand{\rotpages}{rot:epn_toppI}
\AddToHook{shipout/before}{%
  \gdef\vohl@rot{}%
  \@for\vohl@rl:=\rotpages\do{%
    \ifnum\getpagerefnumber{\vohl@rl}=\value{page}\relax\gdef\vohl@rot{/Rotate 90}\fi}%
  \expandafter\pdfpageattr\expandafter{\vohl@rot}}
\makeatother
\usepackage{hyperref}
\hypersetup{
  colorlinks   = true,
  urlcolor     =[RGB]{255, 104, 68},
  linkcolor    =[RGB]{255, 104, 68},
  filecolor    =[RGB]{255, 104, 68},
  citecolor    =[RGB]{255, 104, 68},
  hypertexnames = false
}

\newcommand{\eps}{\varepsilon^{2}}
\makeatletter\newcommand{\tablecaptionhere}{\def\@captype{table}}\makeatother

\newcolumntype{L}[1]{>{\raggedright\arraybackslash}m{#1}}
\defcitealias{toppI}{ToPP~I}

\begin{document}

\title{Distance-based classification of samples of unequal sensitivity can recover the signal-to-noise ratio instead of the physics}
\titlerunning{Sensitivity-driven structure in unsupervised classification}

\author{D.~Vohl\inst{1}\fnmsep\thanks{E-mail: vohl@astron.nl}}
\institute{ASTRON, Netherlands Institute for Radio Astronomy, Oude Hoogeveensedijk 4, 7991 PD Dwingeloo, The Netherlands}
\authorrunning{D.~Vohl}

\date{Received ---; accepted ---}

\abstract
% context
{Distance-based unsupervised classification is applied to astronomical observations of unequal sensitivity, such as pulsar average profiles, where distances reflect both the objects and how well each was observed.}
% aims
{We test whether such a classification groups objects by sensitivity, so that it reports associations with any quantity tied to brightness, and how to detect this.}
% methods
{We apply a standard pipeline (dynamic time warping, DTW; hierarchical clustering; Kruskal--Wallis) to synthetic populations with known truth and to public pulse profiles from the European Pulsar Network (EPN).}
% results
{When morphology is independent of the property tested, the groups sort objects by signal-to-noise ratio (\snr) and so become associated with any quantity correlated with \snr, here a stand-in for the spin down energy loss rate.
Even a weak correlation with \snr yields a significant artefact, whose effect size $\eps$ scales as the square of that correlation.
DTW reduces differences in width but is sensitive to noise, while pointwise distances is near artefact-free.
Comparing groups with \snr flags affected analyses but cannot reject a relation, and neither splitting by brightness nor bootstrap or permutation controls separate artefacts from real relations.
Every EPN sample is sorted by \snr under wide warping ($\eps = 0.26$--$0.85$), and narrowing the \snr range reveals duty-cycle structure, which also links the groups to the spin period and the spin down energy loss rate.
In an earlier EPN classification built on Gaussian-mixture models, the published tree shows no detectable sort and a re-run recovers its complexity ranking but not single edges, while the same distance on observed profiles puts faint pulsars at the tree's edges.}
% conclusions
{Before physical relations are claimed, groups from observations of unequal sensitivity should be compared with \snr and tested with signals injected across a range of noise.}


\maketitle

\section{Introduction}
\label{sec:intro}

Distance-based unsupervised methods promise groups without labels defined in advance.
One chooses a distance measure, or a combination of several, between observations, builds a graph (where objects are vertices joined by edges weighted by their distance) or an embedding (a map in a few dimensions in which the separations between objects approximate their distances) and lets the groups emerge.
We call the division of a sample produced by such a method a partition, and its parts groups.
Distance-based studies have included stellar and quasar spectra~\citep{baron2021}, single pulses of the Vela pulsar~\citep{lousto2022}, catalogue parameters of the pulsar population~\citep{garcia2022}, of fast radio bursts~\citep{chen2022, garcia2024frb} and of millisecond pulsars~\citep{garcia2024msp}, average radio profiles of pulsars~\citep[hereafter \citetalias{toppI}]{toppI}, and high-energy light curves~\citep{garcia2025, torres2026}.
This unsupervised approach does not eliminate assumptions but instead moves them into the choice of distance, and in this context, there is one assumption that is rarely stated. 
That is, the distance between two observations depends only on the difference between the objects themselves.
When working with catalogue spin parameters, this assumption is nearly true given the small errors independent of brightness.
The assumption does not hold however for average profiles, light curves or burst properties measured at unequal sensitivity.

In astronomy, an observation's signal-to-noise ratio (\snr) is set partly by the object through its brightness and distance, and partly by the way it was observed through the telescope, the integration time and instrumental effects. 
Hence, within one sample, \snr can span orders of magnitude, and because faint objects outnumber bright ones, most of the sample lies close to the detection limit.
In pulsar samples, \snr is tied to the physical parameters of interest in several directions.
The spin down energy loss rate, $\dot{E} = 4\pi^{2} I \dot{P}/P^{3}$ (erg\,s$^{-1}$), with $P$ the spin period (s), $\dot{P}$ its time derivative and $I$ the moment of inertia of the neutron star, is largest for young pulsars.
Population models make radio luminosity rise with $\dot{E}$ \citep{arzoumanian2002, fauchergiguere2006}, but young pulsars also lie mostly in the Galactic plane, often at large distances, where scattering and a hotter sky background lower the \snr.
For the same flux density, \snr scales as $\sqrt{(P - W)/W}$ for a pulse of width $W$ \citep{lorimer2004}. 
Given this relation, millisecond pulsars, which are old and have wide pulses relative to their period, are observed at lower \snr than their flux alone suggests.
In heterogeneous archives (i.e. the European Pulsar Network (EPN) database), telescopes, hardware and software backend, integration times and data-reduction pipelines also differ from observation to observation and from pulsar to pulsar.
The sign and size of the correlation of characteristic age, $P/(2\dot{P})$, or $\dot{E}$ with \snr therefore depend on the sample.

The effect of measurement error on clustering, and selection effects causing spurious correlations, are both well documented.
Measurement error degrades clustering, primarily when it is systematic and affects every variable \citep{measurement_error_clustering}. 
Recent work treats error whose size differs between observations (heteroscedastic error) in functional data directly \citep{functional_hetero}.
Benchmarks show that the structure found depends on the choice of distance and algorithm \citep{benchmark_distance}.
Outside astronomy, \citet{cardini2019} show that between-group principal component analysis separates groups clearly even when the groups are drawn at random from the same data.
Selection effects are a well-known source of spurious correlations, in particular when they are combined with scatter, like in the Malmquist bias~\citep{malmquist1922}\footnote{Near a flux limit, only the intrinsically brightest objects enter the sample at large distances, so the sample's mean luminosity is overestimated.} and the Eddington bias~\citep{eddington1913}\footnote{Measurement errors carry more faint objects above a flux limit than bright objects below it, because faint objects are more numerous, so the fluxes measured near the limit are, on average, too high.}.
Differing noise level between objects can create structure that is in line with sensitivity, which in turn a clustering step then converts into an apparently physical association.
The inflation of distances by unequal noise is known in statistics \citep{li2025}, and in single-cell genomics the per-cell detection rate can drive the leading structure \citep{hicks2018}. 
To the best of our knowledge this route to an apparently physical association has not been shown in astronomy.

A variety of distances have been used in unsupervised studies in astronomy.
For instance, a search on the SAO Astrophysics Data System\footnote{ADS, a service that has now merged into Science Explorer (SciX).} search (2026 October) returns 86 astronomy abstracts mentioning dynamic time warping (DTW); in pulsar work it has been used to classify radio profiles \citepalias{toppI}, compare gamma-ray light curves \citep{garcia2025} and X-ray light curves of transitional millisecond pulsars \citep{torres2026}, and follow a magnetar's profile between epochs \citep{rajwade2022}.
None of these pulsar applications tests its distances or groups against \snr, although \citet{torres2026} caution that DTW is harder to apply to a light curve with few counts per bin.

With their Sequencer algorithm, \citet{baron2021} order objects by walking a minimum spanning tree (MST) from its least connected node---with applications to stellar and quasar spectra from the Sloan Digital Sky Survey \citep{york2000}, to seismic dispersion curves \citep{olugboji2017} and to natural images.
Applied to seismograms, the Sequencer revealed waves scattered by previously unrecognised structures near the core--mantle boundary of the Earth \citep{kim2020}.
The algorithm computes four distances between objects (Euclidean, Kullback--Leibler divergence, Earth mover and energy distance), each at several scales obtained by splitting every object into chunks, builds an MST for each distance and scale, and combines them, weighted by the elongation of each tree, into a single ordering.

\citet{baron2021} state their algorithm cannot tell structure from noise and advise to first work with a subset with higher \snr when noise levels vary.
\citetalias{toppI} likewise found, in early tests, that DTW applied directly to some profiles of lower \snr responded to intensity fluctuations, so that noisy profiles appeared similar, and therefore compared Gaussian-mixture models of the profiles instead of the profiles themselves.
Both works anticipate the noise problem without specifically measuring it.

The remainder of this article examines this problem.
While the concepts discussed in this paper are broadly applicable, it focuses on pulsars.
Section~\ref{sec:mechanism} describes how noise inflates distances and sets up test simulations.  
Section~\ref{sec:null} measures the effect, Sect.~\ref{sec:metrics} identifies which distances are vulnerable and why, and Sect.~\ref{sec:diagnostic} assesses diagnostics and controls.
Section~\ref{sec:case} demonstrates the effect with a public collection of pulse profiles, including models from \citetalias{toppI}.
Finally, a discussion is presented in Sect.~\ref{sec:discussion} before concluding in Sect.~\ref{sec:conclusions}.

\section{Noise inflation and simulated populations}
\label{sec:mechanism}

Let an object $i$ be observed as $x_{i} = s_{i} + e_{i}$ over $M$ samples, where $s_{i}$ is the true signal and $e_{i}$ is independent zero-mean noise of scale $\sigma_{i}$.
For two objects $i$ and $j$, the squared Euclidean distance $d^{2}(x_{i}, x_{j})$ is the sum over the $M$ samples of the squared difference between $x_{i}$ and $x_{j}$, and its expected value $\mathbb{E}$ over noise realisations is
\begin{equation}
\mathbb{E}\big[d^{2}(x_{i}, x_{j})\big]
   \;=\; d^{2}(s_{i}, s_{j}) \;+\; M\,(\sigma_{i}^{2} + \sigma_{j}^{2}).
\label{eq:inflation}
\end{equation}
Noise therefore inflates every squared distance from object $i$ by the same amount, $M\,\sigma_{i}^{2}$.
In this scenario, a noisy object is pushed away from all other objects, including those with the same inherent (but less noisy) signal.
Equation~(\ref{eq:inflation}) holds only if the noise of different objects is independent.
Noise shared by objects $i$ and $j$, such as interference or a calibration error common to one observing session, adds the term $-2\sum_{m=1}^{M}\mathrm{Cov}(e_{im}, e_{jm})$, where $e_{im}$ is the noise of object $i$ in sample $m$ and $\mathrm{Cov}$ is the covariance, so positively correlated noise reduces the distance between those objects and can draw them together.

Noise moves each object in a random direction, but in $M$ dimensions these directions are nearly orthogonal to one another and to the separations between objects, so every distance grows and each faint object is pushed outwards, away from the rest of the sample. 
We refer to this expulsion as a radial displacement.
A low \snr object ends up further from every other object, including other faint ones, and ends up sitting at the edge of the graph or embedding.
Measurement quality becomes the main basis of the grouping, and the partition inherits an association with any quantity relating to brightness.
Unlike the Malmquist and Eddington biases which distort quantities measured from flux, this bias distorts the structure of a graph built from distances between shapes.
It is closer to the noise bias of weak-lensing shape measurements, which grows as the \snr of each galaxy falls \citep{refregier2012}.
No flux is used, and brightness enters only through the noise level of each observation.
We show in the following Sections that every distance is inflated in this way, including the Euclidean distance.

\subsection{Simulated populations}
\label{sec:populations}

We test if noise inflation alone can produce groups that are ordered by brightness. 
We proceed by using two simulated populations in which we know which properties change the profile shapes and which are tied only to \snr.
The null population has $n = 180$ objects, whose profile have $M = 256$ phase samples and $1$ to $3$ Gaussian components of random amplitude and separation ($8$ to $22$ samples) and fixed width ($w = 6$ samples, the standard deviation of each Gaussian), with \snr log-uniform over $8$ to $400$.
Each object also carries a property $y$, standing in for $\log\dot{E}$ (Sect.~\ref{sec:intro}).
The property is only tied to $\log\snr$ through $y = 0.9\,z + \mathcal{N}(0, 0.9^{2})$, where $z$ is the standardised $\log\snr$ and $\mathcal{N}(0, 0.9^{2})$ is a normal random variable with mean zero and variance $0.9^{2}$.
This gives $\rho(y, \log\snr) \simeq 0.71$, where $\rho$ is the Spearman rank correlation.
Each noise-free profile has a peak of $1$, white Gaussian noise of standard deviation $\sigma_{i} = 1/\snr_{i}$ is added independently to each sample, and the observation is then normalised to its peak.

The two-property population is built in the same manner, only with two properties instead of one.
A genuine property is drawn from a standard normal distribution independently of \snr.
It is scaled linearly to $8$--$22$ samples and sets the separation of the components.
A spurious property is tied to $\log\snr$ exactly as $y$ is in the null population, and to nothing else.
Section~\ref{sec:diagnostic} also uses a variant in which the genuine property is tied to \snr, and Appendix~\ref{app:supp_diag} two further ones, each described where it is used.

\subsection{Pipeline}
\label{sec:pipeline}

DTW matches each sample of one profile to a sample of the other, allowing the match to shift in phase within a band, so that profiles differing in width or component spacing can still be aligned before they are compared.
The pipeline is pairwise DTW (symmetric step pattern, absolute local cost, Sakoe--Chiba band of half-width $38$ of $256$ samples, divided by the summed length of the two sequences; \citealt{sakoe1978}), followed by hierarchical clustering (Ward linkage into $k = 5$ groups unless stated otherwise).
The band of $0.15$ rotations is narrow enough that the warping cannot match a main pulse to an interpulse, which lie about half a rotation apart, as intended in \citetalias{toppI}; how the results depend on the band is shown in Sect.~\ref{sec:metrics}.
The association between the groups and a property is measured by the Kruskal--Wallis \citep{kruskal1952} effect size $\eps = (H - k + 1)/(n - k)$, with $H$ the Kruskal--Wallis statistic, $k$ the number of groups and $n$ the number of objects, which is the fraction of rank variance explained by the groups \citep[$\eta^{2}_{H}$ in][]{tomczak2014}.
It is set to zero when negative, and groups of fewer than three objects are removed.
We write $\eps(x)$ for the effect size of a partition against a variable $x$, such as $\eps(y)$ or $\eps(\log\snr)$.
Ward linkage on a dissimilarity that is not Euclidean, such as DTW, no longer minimises the variance within groups \citep{murtagh2014}; we use it because it is the common default.

\section{Noise alone produces associations}
\label{sec:null}

\begin{figure*}
\centering
\includegraphics[width=\textwidth]{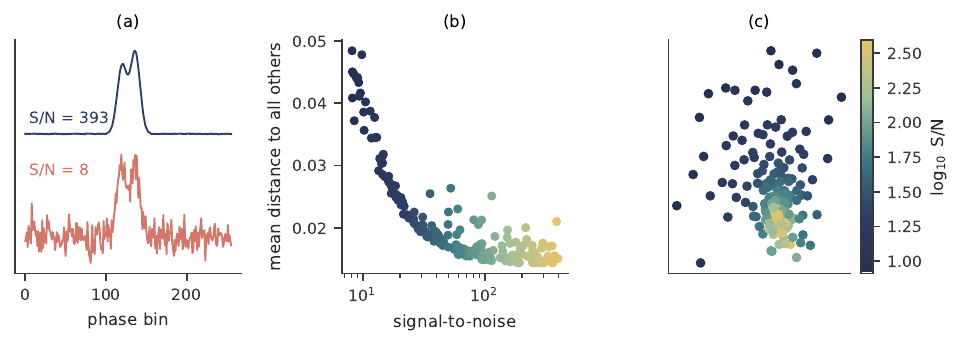}
\caption{One realisation of the synthetic two-property population of Sect.~\ref{sec:populations}, in which morphology is set by a property independent of brightness.
\textbf{(a)} Two objects of the same kind observed at \snr 393 and 8.
\textbf{(b)} An object's mean distance to all others rises as its brightness falls (Spearman rank correlation $\rho = -0.86$).
\textbf{(c)} In a two-dimensional metric multidimensional scaling embedding of the distances, whose axes have no units, well-observed objects sit in the centre and poorly observed objects at the edge.
The null population behaves in the same way.}
\label{fig:mechanism}
\end{figure*}

To verify Eq.~(\ref{eq:inflation}), we draw ten pairs of profiles from the null population and add noise to both profiles of each pair at \snr $6$ to $200$, with $200$ noise realisations at each level.
The excess grows as $\sigma^{2.00}$ and agrees with Eq.~(\ref{eq:inflation}) to within $1$ per cent at every level.
Most DTW implementations instead use the absolute difference $|a - b|$ between matched samples $a$ and $b$ as local cost.
The excess then grows as $\sigma^{1.00}$ to $\sigma^{1.08}$.
On the two-property population, the mean distance of object to all others rises as its brightness falls, and in an embedding of the distances the well-observed objects sit in the centre and the poorly observed ones at the edge (Fig.~\ref{fig:mechanism}).

In the null population this displacement becomes a spurious association.
We define such an association---which originates from \snr alone and not from morphology---an artefact, and measure its size by $\eps(y)$.
Other clustering algorithms, compared in full in Sect.~\ref{sec:metrics}, give significant associations as well as Ward linkage.
A list of every DTW configuration that splits the sample into real groups instead of into one large group plus singletons is presented in Table~\ref{tab:null}.
Morphology does not depend on $y$ in this population, and yet every configuration leads to an association, with $\eps(y)$ from $0.05$ to $0.46$ and $p$ as low as $10^{-18}$.
On real data, such values would likely be interpreted as evidence of a physical relation.
These values come from one realisation (Appendix~\ref{app:seeds}).
Over ten realisations, the Ward partition with five groups gives $\eps(y)$ between $0.25$ and $0.45$. 
We therefore only draw conclusions from comparisons between methods that hold in every realisation.
The mean distance from an object to all others correlates with $\log\snr$ at $\rho = -0.88$ and with $y$ at $-0.67$.

\begin{table}
\centering
\caption{Association between morphology and $y$ in a population where none exists (DTW distances, one realisation; every configuration that partitions).
$k'$ is the number of groups with at least three members.
Average linkage and the MST degenerate into one large group plus singletons and are omitted.}
\label{tab:null}
\begin{tabular}{lrrrrr}
\toprule
algorithm & $k$ & $k'$ & $\eps(y)$ & $p$ & $\eps(\log \snr)$ \\
\midrule
ward       & 3 & 3 & 0.337 & $4\times10^{-14}$ & 0.618 \\
       & 5 & 4 & 0.335 & $3\times10^{-13}$ & 0.604 \\
       & 8 & 5 & 0.325 & $3\times10^{-12}$ & 0.595 \\
\midrule
complete   & 3 & 3 & 0.054 & $3\times10^{-3}$  & 0.140 \\
   & 5 & 3 & 0.106 & $3\times10^{-5}$  & 0.231 \\
   & 8 & 3 & 0.085 & $3\times10^{-4}$  & 0.206 \\
\midrule
$k$-medoids& 3 & 2 & 0.272 & $2\times10^{-12}$ & 0.535 \\
& 5 & 4 & 0.085 & $5\times10^{-4}$  & 0.129 \\
& 8 & 8 & 0.419 & $2\times10^{-14}$ & 0.719 \\
\midrule
spectral   & 3 & 3 & 0.448 & $2\times10^{-18}$ & 0.805 \\
   & 5 & 5 & 0.396 & $5\times10^{-15}$ & 0.699 \\
   & 8 & 8 & 0.463 & $6\times10^{-16}$ & 0.766 \\
\bottomrule
\end{tabular}
\end{table}

\begin{figure}
\centering
\includegraphics[width=\columnwidth]{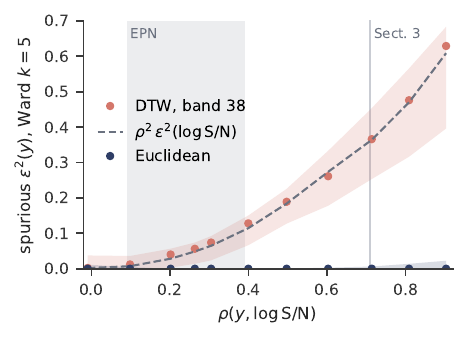}
\caption{Spurious association against the property's tie to \snr, growing quadratically under DTW (null population, DTW and Euclidean distance, Ward, $k = 5$; median and full range of ten realisations at each coupling).
The dashed line is $\rho^{2}\,\eps(\log\snr)$, measured on the same DTW partitions.
The grey band is the range of $\rho(\log\dot{E}, \log\snr)$ in the samples of Sect.~\ref{sec:case}; the vertical line marks the coupling used elsewhere in this paper.}
\label{fig:coupling}
\end{figure}

As we will see in Sect.~\ref{sec:case}, this coupling of $y$ to \snr is stronger than in real pulse profiles.
The values in the rest of this paper, at $\rho = 0.71$, are therefore close to the worst case for pulsars.
The effect size scales quadratically with the coupling, $\rho(y, \log\snr)$ (Fig.~\ref{fig:coupling}).
Over ten realisations at each of eleven couplings, the DTW partition gives $\eps(y) \simeq \rho^{2}\,\eps(\log\snr)$ (rank correlation $0.98$ between the two sides over $300$ partitions at $k \in \{3, 5, 8\}$; median ratio $1.03$).
Both factors can be measured on the data: $\rho$ from the property and \snr alone, and $\eps(\log\snr)$ from the partition.
The product ($\rho^{2}\,\eps(\log\snr)$) estimates the association that would be produced by brightness alone.
When $\eps(y)$ is close to this product, it can therefore be explained by brightness. 
An excess above this product hints at a correlation with morphology (at least for populations and pipelines similar to ours).
Under brightness alone the excess reaches $0.04$ to $0.06$ (95th percentile, $180$ objects). 
Real relations in our simulations exceed in only $19$ to $31$ of $100$. 
Hence, a large excess supports a real relation but a small one does not rule it out.
At $\rho = 0.26$ the artefact is $\eps(y) = 0.06$ (median; $0.01$--$0.07$) and significant at $p < 0.01$ in five of ten realisations, at $\rho = 0.30$ it is $0.07$ and significant in nine, and at $\rho = 0.10$ it is significant in none.
Euclidean distance results in $\eps(y) \leq 0.02$ for every coupling. 
Hence, with the same clustering, when DTW is replaced by a non-warping distance, the artefact almost vanishes.

The effect is driven by the range of sensitivity.
Narrowing the \snr range from $[8, 400]$ to $[70, 80]$ in four steps weakens the correlation of distance with \snr steadily, from $-0.87$ to $-0.19$ (medians of ten realisations); at a factor of three the partition carries the artefact in six of ten realisations ($\eps(y) \in [0.14, 0.28]$) and not in the other four, and does not occur below a factor of two.
It is not a small-sample problem nor due to the simulated \snr range, since from $n = 60$ to $480$ the effect size stays at $0.35$--$0.39$ while $p$ falls from $3\times10^{-5}$ to $1\times10^{-39}$, and widening \snr to the $5$--$1100$ of the EPN samples gives $0.34$ (medians of ten realisations).

\section{Distance measures and their sensitivity to noise}
\label{sec:metrics}

Using identical observations, we test the sensitivity of distance measures to noise by crossing four distances with six clustering algorithms, and three values of the number of groups, $k = 3$, $5$ and $8$.
The distance measures are DTW, Euclidean distance, the $L_{1}$ distance (the sum over samples of the absolute differences) and correlation distance (one minus the Pearson correlation between two profiles).
The clustering algorithms are three forms of hierarchical clustering, $k$-medoids, spectral clustering and a cut of the MST.
Hierarchical clustering merges, at each step, the two groups whose union least increases the variance within groups (Ward linkage; \citealt{ward1963}), whose most distant members are closest (complete linkage) or whose members are closest on average (average linkage).
$k$-medoids assigns each object to the nearest of $k$ representative objects, chosen among the objects themselves \citep{kaufman1990}.
Spectral clustering converts the distances into a Gaussian affinity, with scale set by the median distance; it groups objects by the leading eigenvectors of its normalised graph Laplacian \citep{ng2002, vonluxburg2007}.
The MST cut, a simplified form of the method of \citet{zahn1971}, removes the $k-1$ longest edges of the tree, and groups each connected piece remaining.

We find the artefact to be only severe for DTW.
Under Ward linkage at $k = 5$, DTW gives $\eps(y) = 0.25$--$0.45$ over ten realisations (median $0.37$), while the other three give $\eps(y) \leq 0.03$.

\begin{figure*}
\centering
\includegraphics[width=\textwidth]{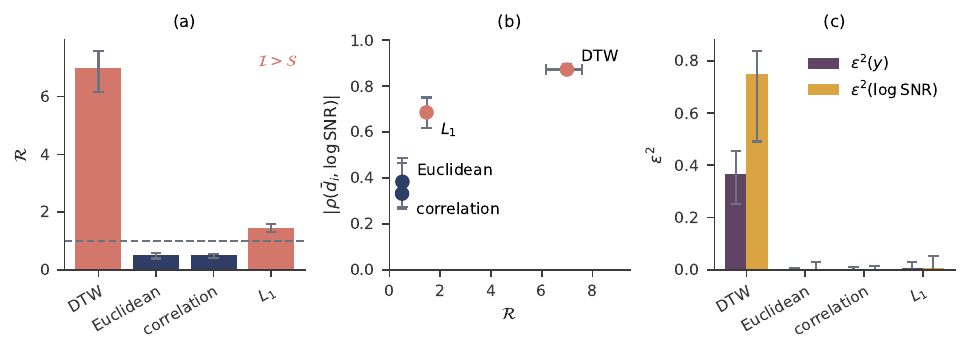}
\caption{Sensitivity of four distance measures to noise (null population, Ward, $k = 5$; median and full range of ten realisations).
\textbf{(a)} Noise-to-structure ratio $\mathcal{R}$ for four distances on identical observations.
\textbf{(b)}  Correlation of mean distance with $\log\snr$ against $\mathcal{R}$ (coloured as in (a)); $\mathcal{R}$ tracks the radial displacement.
\textbf{(c)} Effect sizes of the Ward partition against $y$ and $\log\snr$; only DTW yields a spurious association under Ward linkage.}
\label{fig:metrics}
\end{figure*}

Given clustering acts on the contrast between distances, we define the noise-to-structure ratio
\begin{equation}
\mathcal{R} \;=\; \frac{\mathcal{I}}{\mathcal{S}} \;=\;
\frac{\Delta_{5\text{--}95}\big[\,\bar{d}_{i}(\mathrm{observed}) -
      \bar{d}_{i}(\mathrm{true})\,\big]}
     {\mathrm{sd}\big[\,d_{ij}(\mathrm{true})\,\big]},
\label{eq:ratio}
\end{equation}
where $\bar{d}_{i}$ is the mean distance from object $i$ to all others, computed on both the observed (noisy) and the true (noise-free) profiles, $\Delta_{5\text{--}95}$ is the range between the 5th and 95th percentiles over objects, and $\mathrm{sd}$ is the standard deviation over all pairs of the true distances $d_{ij}$.
Thus, $\mathcal{R}$ is the ratio of the noise inflation $\mathcal{I}$, the spread over objects of the excess in their mean distance, to the spread $\mathcal{S}$ of the true pairwise distances.
On simulated data, $\mathcal{R}$ ranges from $6.2$ to $7.6$ for DTW over ten realisations, $1.3$ to $1.6$ for $L_{1}$ and below $0.6$ for Euclidean and correlation distance---tracking the radial displacement (Fig.~\ref{fig:metrics}).
$\mathcal{R}$ separates wide warping from the other distances, but it does not order moderate configurations.
For instance, $L_{1}$ has $\mathcal{R} = 1.46$ and $\eps(y) = 0.008$, while DTW with a half-width of $4$ has $\mathcal{R} = 0.90$ and $\eps(y) = 0.087$ (Table~\ref{tab:band}).
Nor does $\mathcal{R}$ predict the size of the artefact, as it does not indicate how strongly the property follows brightness (Appendix~\ref{app:supp_metrics}).

The cause of the vulnerability is not the normalisation but the warping procedure.
For profiles of equal length, the divisor is the same for every pair. 
When removed, $\mathcal{R}$ and the partition remain unchanged, and DTW with a band of zero width equals the $L_{1}$ distance divided by $M$.
As the band widens (Table~\ref{tab:band}), the spread of the true distances falls to a tenth of its unwarped value while the noise inflation halves between half-widths of $0$ and $4$ and then barely changes.
Whereas a pointwise distance compares sample $x_{i}$ of one profile with sample $y_{i}$ of the other, DTW also considers neighbouring samples within the band, allowing for shifts and stretches in phase, so that differences in width count for less than in a pointwise comparison, although they are not removed \citepalias{toppI}.
Warping cannot, however, align independent noise.

On the two-property population (Sect.~\ref{sec:populations}), we find that widening the band moves the partition from the genuine relation to the artefact.
Component separation is close to profile width, which warping is often used to ignore. We therefore also coded the genuine relation in a property that does not depend on width---the amplitude ratio of two components---with each profile stretched to a width drawn independently of everything else.
Wide warping loses this relation too ($\eps = 0.55$ with no band, $0.03$ at half-width $38$; Table~\ref{tab:band}), while bands of up to $8$ samples retain it ($0.49$--$0.51$) with almost no artefact ($\eps \leq 0.02$).
Furthermore, normalising by the realised path length ($\eps(y) = 0.34$) or using a squared local cost ($0.33$) does not help.

The synthetic profiles are centred on their true signal, which favours the pointwise distances.
Centring them instead on their noisy peak, as for real profiles, creates no artefact for the pointwise distances but costs them part of the genuine relation, which the peak no longer locates reliably (Appendix~\ref{app:supp_metrics}).

\begin{table*}
\centering
\caption{Effect of the DTW band half-width, in samples, with $\mathcal{R} = \mathcal{I}/\mathcal{S}$ (Eq.~(\ref{eq:ratio})).
Values are medians over ten realisations, and $\eps$ is that of the Ward partition.
A half-width of zero is the $L_{1}$ distance.}
\label{tab:band}
\footnotesize
\setlength{\tabcolsep}{4pt}
\begin{tabular}{lrrrrrrrr}
\toprule
 & \multicolumn{4}{c}{null population} & \multicolumn{2}{c}{two-property} & \multicolumn{2}{c}{amplitude ratio} \\
half-width & $\mathcal{R}$ & $\mathcal{I}$ & $\mathcal{S}$ & $\eps(y)$ & genuine & spurious & genuine & spurious \\
\midrule
0   & 1.46 & 0.047 & 0.032  & 0.008 & 0.28 & 0.00 & 0.55 & 0.00 \\
4   & 0.90 & 0.023 & 0.026  & 0.087 & 0.11 & 0.07 & 0.49 & 0.00 \\
8   & 1.21 & 0.024 & 0.019  & 0.217 & 0.19 & 0.18 & 0.51 & 0.01 \\
16  & 3.15 & 0.024 & 0.0077 & 0.323 & 0.12 & 0.40 & 0.37 & 0.19 \\
38  & 6.99 & 0.024 & 0.0034 & 0.366 & 0.03 & 0.42 & 0.03 & 0.34 \\
\midrule
Euclidean & 0.51 & & & 0.000 & 0.22 & 0.00 & 0.55 & 0.00 \\
Euclidean, common width & & & & 0.000 & 0.14 & 0.00 & 0.69 & 0.00 \\
\bottomrule
\end{tabular}
\end{table*}

While the distance sets the noise inflation, the clustering algorithm determines how much of it becomes groups.
Here, spectral clustering is the most vulnerable: combined with DTW, it returns the largest artefact of our set ($\eps(y) = 0.463$). 
Furthermore, it also produces one with $L_{1}$, Euclidean and correlation distance ($0.06$--$0.20$ at $k = 8$) in this population.
Average linkage and the MST cut do not partition this sample (Table~\ref{tab:null}).

The usual ways of handling noise in DTW, smoothing and derivative warping, do not remove the artefact (Table~\ref{tab:remedies}).
Suppressing the off-pulse region does not remove it either, since clipping the baseline as in the mixture models of \citetalias{toppI} roughly halves the association with the property tied to brightness ($\eps = 0.24$ against $0.45$), and keeping an on-pulse window reduces it to about a third ($0.14$; Table~\ref{tab:remedies}).
On the two-property population, no combination of smoothing and derivative warping (DTW applied to the derivative of each profile) recovers the genuine relation while removing the spurious association.
Smoothing over ten samples comes nearest to removing the artefact, but only by erasing the morphology, in turn not recovering the genuine relation either. 
Derivative warping on unsmoothed profiles is the worst, since differentiation amplifies the noise, which hides the morphology but leaves the artefact intact.

The best alternative we found removes width explicitly instead of by warping.
By stretching each profile about its peak, and having its half-maximum width falling on a common number of samples, and then performing a pointwise comparison keeps the asymmetry relation ($\eps = 0.69$, or $0.68$ with widths measured on the noisy profiles) and produces no artefact in the null or two-property populations ($\eps(\log\snr) \leq 0.04$). 
DTW on the same stretched profiles still sorts them by \snr ($\eps = 0.59$--$0.70$).
The pointwise distance is not free of noise, since an object's mean distance still follows \snr ($\rho = -0.29$, against $-0.79$ for DTW on the same profiles), but the inflation is too small to form groups.
Defining the width by the outermost half-maximum crossings, or centring profiles on their noisy peak, still favours the stretched comparison over DTW (Appendix~\ref{app:supp_metrics}).
It appears that narrow bands are safe only sometimes.
At $8$ samples ($3$ per cent of the profile) the artefact is $0.01$ in the asymmetry population but $0.18$ in the two-property population (Table~\ref{tab:band}), and a band of $13$ samples ($5$ per cent) gives $\eps(y) = 0.31$ in the null population (median of five realisations), so the safe width depends on the profiles and has to be checked by simulation.

\section{Diagnostics and controls}
\label{sec:diagnostic}

As ground truth is unknown in real data, this section evaluates four common safeguard checks that can be applied to real data and tests whether they can disentangle an artefact from a genuine relation.

\subsection{Comparing the partition with \snr}
\label{sec:power}

The simplest check compares the partition's effect size against \snr, $\eps(\log\snr)$, with its effect size against the property, $\eps(y)$, as is done for batch effects in genomic data \citep{leek2010}.
The check flags a sensitivity sort when $\eps(\log\snr) \geq \eps(y)$.
In Table~\ref{tab:null}, $\eps(\log\snr)$ always exceeds $\eps(y)$.
We apply the check to the null population, to the two-property population, and to a tied variant of it, in which the genuine property is tied to $\log\snr$ exactly as $y$ is in the null population, with $24$ realisations of each ($180$ objects at $256$ samples, as in Table~\ref{tab:null}).
The check flags a sensitivity sort in all $72$ realisations, whether the genuine property is tied to brightness (median $\rho(y,\log\snr)=0.71$) or not (median $-0.01$).
Where it is not tied, the DTW partition carries almost none of the genuine relation (median $\eps(y) = 0.01$).
On $L_{1}$ and Euclidean distance, it flags a sort in at most $3$ of $100$ realisations of our strong real relations, but in $13$ to $50$ per cent of the significant realisations of a weaker one tied to \snr.
Under DTW, it flags one in nearly all realisations, so it is not a test of soundness and cannot reject a relation.
DTW being a distance measure particularly sensitive to noise, we observe that the partition follows brightness whether or not a real relation exists.

\subsection{Splitting by brightness}
\label{sec:strat}

A spurious association anchored on sensitivity should weaken in the brighter part of the sample, while a real relation should hold or strengthen.
We measure the relation by the rank correlation of each object's mean distance with the property, and take it as a real morphology–property relation when this correlation, in absolute value, is larger in the brightest quarter of the sample than in the faintest half.

We test this check on three kinds of relation\footnote{Two more kinds are discussed in Appendix~\ref{app:supp_diag}.}, $300$ realisations of each (Table~\ref{tab:strat_mech}; Fig.~\ref{fig:diagnostic}b shows one realisation).
Here, a pure artefact corresponds to the null population, and a real relation independent of brightness is the genuine property of the two-property population.
A real relation tied to brightness is the tied variant of Sect.~\ref{sec:power}.
A false positive is an artefact that the check reads as real, and a false negative is a real relation that it reads as an artefact.
The check gives almost no false positives on a pure artefact ($1$ per cent), and false negatives on a real relation independent of brightness in $11$ per cent of realisations.
Unfortunately, in the realistic case for $\dot{E}$, where a real relation is also tied to brightness, it gives false negatives in $74$ per cent.

We also test a weaker coupling, $\rho(y,\log\snr) = 0.30$, closer to the empirical samples of Sect.~\ref{sec:case}.
False negatives on a real relation tied to brightness drop to $10$\%, but a pure artefact gives false positives in $45$\%.
Requiring $|\rho|$ in the brightest quarter to exceed that in the faintest half by $0.15$, one standard error of $\rho$ for its $45$ objects, cuts the false positives to $7$\% and raises the false negatives to $34$\%, but the margin does not carry over to larger samples, since for $360$ objects it misses every real relation tied to brightness at $\rho = 0.71$ (Appendix~\ref{app:supp_diag}).

These rates hold for the correlation of the mean distance with the property; for another statistic they have to be measured again.
Increasing our reading of the correlation over five brightness bins does not help, because both a pure artefact and a real relation tied to brightness give a correlation that is positive in some bins and negative in others  (Appendix~\ref{app:supp_diag}).
As a check, splitting by brightness can reveal an artefact.
However, it cannot provide a definitive verdict.

Holding \snr fixed, by shuffling the property among objects of similar \snr and comparing it with the mean DTW distance, rarely passes an artefact of brightness (about $2$ in $100$) and finds $89$ to $95$ in $100$ real relations tied to brightness (Table~\ref{tab:strat_mech}).
No such test can tell a real relation from a selection on \snr, which it reports as real in $79$ of $100$. 
Hence, passing the test only means that the association cannot be explained by brightness alone.

\subsection{Removing the noise contribution}
\label{sec:correction}

When the contribution of noise to a statistic cannot be written down as a formula---as for DTW, whose warping path is itself chosen on the noisy data---it can be estimated by simulation, as is done in the calibration of noise bias in weak-lensing shape measurements \citep{kacprzak2012}.
For each object, we subtract from the measured statistic its median over eight copies of a noise-free model with noise added at the object's measured level.
We test this on injected profiles of one or two components at \snr\ $15$ to $200$, with the mean absolute residual from a single-Gaussian fit as the statistic.
When uncorrected, the same profile scores $2.4$ to $13$ times higher at \snr\ $15$ than at $200$, while  when corrected it stays below $0.001$ for one component and varies by $9$--$19$ per cent across \snr\ for two.
Subtracting only the noise part of this floor corrects a statistic averaged over many objects, but not the partition, which is built from individual noisy distances and still follows \snr ($\eps(\log\snr) = 0.21$) after the expected noise, computed from the true profiles, is subtracted from every distance (Appendix~\ref{app:supp_null}).

\subsection{Stable but not real}
\label{sec:controls}

We ran the bootstrap and fresh-noise resampling used to assess the stability of clusters \citep{hennig2007} and a permutation test---three common controls---on the null population (ten realisations, DTW, Ward, $k = 5$).

The bootstrap over objects reproduces the partition at a median adjusted Rand index (the agreement between two partitions, $1$ for identical ones and $0$ for chance) of $0.64$ ($0.50$--$0.91$ over the ten realisations). 
The fresh noise realisation at each object's own level reproduces it at $0.66$ ($0.49$--$0.88$). 
Finally, permuting the property gives $p = 0.001$ in all ten realisations, the smallest value $999$ permutations allow.

The spurious association passes all three controls as stable and significant.
All controls test if the analysis is self-consistent, and it is.
Each resample and noise draw keeps the same sensitivity distribution, recreating the displacement that made the structure. 
Finally, a permutation test cannot tell apart an association through brightness from a genuine physical association.

In essence, injecting ground truth at one noise level ends up fixing the variable that needs to vary.
A new morphology statistic should be tested by injecting the same signal at a wide range of noise levels and checking that the recovered value stays the same.

\section{Public pulse profiles}
\label{sec:case}

In this section, we evaluate the extent of the artefact on empirical pulsar data. 

\subsection{Data}

We applied the pipeline to the European Pulsar Network database of published profiles\footnote{Following the procedure of \citetalias{toppI}, we retrieved profiles from \url{https://psrweb.jb.man.ac.uk/epndb/} on October 14, 2026, for a total of $7\,082$ profiles.}, the data set classified by \citetalias{toppI}.
Profiles are split into the four frequency bands of that work ($400$--$700$, $700$--$1000$, $1000$--$1500$ and $1500$--$2000$\,MHz), and each pulsar contributes at most one profile per band.
Unlike \citetalias{toppI}, which kept the brightest profile in each band, required $\snr > 20$ and re-gridded to $512$ bins, we keep the profile closest to the band centre, apply only a floor of $\snr \geq 5$, and rebin to $256$ bins.

This sampling matches the simulations and is close to the native resolution of many EPN profiles, a third of which have fewer than $512$ bins in the largest sample\footnote{Appendix~\ref{app:supp_case} repeats the analysis at $512$ samples.}.
Profiles are then peak-centred, median-subtracted and peak-normalised.
We measure \snr with the same estimator, which divides the peak minus the median of the medians of $8$-sample chunks by the median of their standard deviations, here on the $256$-sample profile.

On the public profiles, the property tested against morphology is $\log\dot{E}$ itself, for which $y$ stood in the simulations (Sect.~\ref{sec:populations}), with $\dot{E}$ taken from the \texttt{EDOT} column of the ATNF catalogue \citep[version 2.8.1;][]{atnf}, which assumes $I = 10^{45}$\,g\,cm$^{2}$.
Profiles of pulsars without a catalogue $\dot{E}$ are left out, at most $33$ in any sample.
The positive control is the duty cycle $W_{50}/P$, with $W_{50}$ (ms) the pulse width at half maximum from the \texttt{W50} column of the same catalogue, 40 to 50 per cent of which, depending on the band, come from observations at about $1.3$\,GHz \citep{posselt2021}, and the rest from other publications; since profile widths change with frequency, this control is weakest for the $606/610$\,MHz sample.
The catalogue flux density at 1400\,MHz, $S_{1400}$ (mJy), is an independent measure of brightness.

Four samples combine all publications, one per band, and three come from single publications, to separate telescopes and pipelines from \snr: \citet{jk18} at $1369$\,MHz, and \citet{gl98} at $1408$ and at $606/610$\,MHz (Table~\ref{tab:case_observed}).\footnote{The EPN database labels \citet{jk18} as \texttt{jk17}, with the year 2017, and our code and data products follow that label.}
The samples overlap, since \citet{gl98} provide most of the profiles at $400$--$700$ ($273$ of $485$), $700$--$1000$ ($215$ of $263$) and $1500$--$2000$\,MHz ($232$ of $327$), and \citet{jk18} provide $572$ of the $1359$ at $1000$--$1500$\,MHz, with \citet{gl98} another $97$, so the seven samples rest largely on two publications.
In each sample, \snr spans about two decades, from $7$--$12$ to $259$--$1131$ between the 5th and 95th percentiles.

\subsection{As observed, every sample is sorted by \snr}

\begin{table*}
\centering
\caption{The pipeline on EPN profiles as observed (DTW, Ward).
$\bar{d}$ is a profile's mean distance to all others; $n$ is the number of profiles.
Effect sizes are for $k = 5$, with the range over $k = 3$, $5$ and $8$ in parentheses.
$\rho_{\dot{E}} = \rho(\log\dot{E}, \log\snr)$, and $\rho_{\dot{E}}^{2}\,\eps(\log\snr)$ is the part of $\eps(\log\dot{E})$ expected from brightness alone (Sect.~\ref{sec:null}).}
\label{tab:case_observed}
\setlength{\tabcolsep}{3pt}
\small
\begin{tabular}{lrrrrrrrr}
\toprule
sample & $n$ & \multicolumn{2}{c}{$\rho(\bar{d}, \cdot)$} & \multicolumn{3}{c}{$\eps(\cdot)$} & $\rho_{\dot{E}}$ & $\rho_{\dot{E}}^{2}\,\times$ \\
\cmidrule(lr){3-4} \cmidrule(lr){5-7}
 & & $\log\snr$ & $\log S_{1400}$ & $\log\snr$ & $\log\dot{E}$ & duty & & $\eps(\log\snr)$ \\
\midrule
all papers, 400--700\,MHz    & 485  & $-0.82$ & $-0.34$ & 0.26 (0.26--0.58) & 0.02 (0.02--0.03) & 0.07 (0.06--0.09) & 0.12 & 0.00 \\
all papers, 700--1000\,MHz   & 263  & $-0.83$ & $-0.57$ & 0.85 (0.80--0.85) & 0.04 (0.04--0.04) & 0.12 (0.03--0.12) & 0.23 & 0.05 \\
all papers, 1000--1500\,MHz  & 1359 & $-0.78$ & $-0.39$ & 0.54 (0.54--0.56) & 0.02 (0.02--0.06) & 0.09 (0.09--0.26) & 0.14 & 0.01 \\
all papers, 1500--2000\,MHz  & 327  & $-0.81$ & $-0.55$ & 0.67 (0.66--0.86) & 0.09 (0.09--0.12) & 0.05 (0.02--0.05) & 0.39 & 0.10 \\
\citet{jk18}, 1369\,MHz      & 594  & $-0.88$ & $-0.59$ & 0.64 (0.63--0.64) & 0.01 (0.01--0.01) & 0.17 (0.17--0.23) & 0.09 & 0.01 \\
\citet{gl98}, 1408\,MHz      & 275  & $-0.83$ & $-0.54$ & 0.81 (0.44--0.81) & 0.09 (0.05--0.09) & 0.05 (0.05--0.06) & 0.26 & 0.05 \\
\citet{gl98}, 606/610\,MHz   & 276  & $-0.85$ & $-0.43$ & 0.57 (0.23--0.57) & 0.03 (0.02--0.03) & 0.06 (0.04--0.06) & 0.19 & 0.02 \\
\bottomrule
\end{tabular}
\end{table*}

With five groups, the partition explains $26$ to $85$ per cent of the rank variance of \snr across the seven samples, against at most $9$ per cent for $\dot{E}$ and $17$ per cent for the duty cycle (Table~\ref{tab:case_observed}).
The lowest value is at $400$--$700$\,MHz, where one group holds $427$ of the $485$ profiles; with eight groups it rises to $58$ per cent.
The mean distance follows \snr just as for simulated data, and the catalogue flux density. 
The sort is a reflection of the quality of the measurement, which is set partly by the pulsar's brightness and partly by the observation setup and post-processing.
\snr also rises with the native number of bins in the two largest samples ($\rho = 0.43$ and $0.46$).
The \snr sort is just as strong in the single-publication samples, and the partition follows the publication a profile came from, its native bin count, dispersion measure, catalogue scattering time and observing frequency only weakly (Appendix~\ref{app:supp_case}).
Because $\dot{E}$ is only weakly tied to \snr in this database ($\rho = 0.09$ to $0.39$), its association with the partition is small, and splitting by brightness cannot show whether any of it is real (Sect.~\ref{sec:strat}).

A pointwise distance on the same profiles does not sort them by \snr but by width (Table~\ref{tab:case_width}).
Euclidean, $L_{1}$ and correlation distances give $\eps(\log\snr) = 0.00$--$0.15$ and $\eps(\mathrm{duty}) = 0.18$--$0.61$.
That is what a distance that compares overlapping profiles should do. 
However, this is not what \citetalias{toppI} aimed to measure: it was the average profile shape abstracted from width.
We therefore stretched each profile so that its half-maximum width, measured directly on the profile, falls on $24$ samples. 
The pointwise partitions of the stretched profiles follow neither \snr ($\eps \leq 0.05$) nor $\dot{E}$ ($\leq 0.07$), while DTW on the same stretched profiles still sorts them by \snr ($0.22$--$0.51$).
Thus, in these profiles the sort comes from the warping and not from the removal of width.
Since stretching also removes the duty cycle, representing our only positive control ($\eps \leq 0.08$), we cannot show here what shape structure the stretched comparison recovers, only that it is not a sort by brightness.

\subsection{Narrowing the \snr range reveals shape}
\label{sec:case_narrow}

\begin{figure}
\centering
\includegraphics[width=\columnwidth]{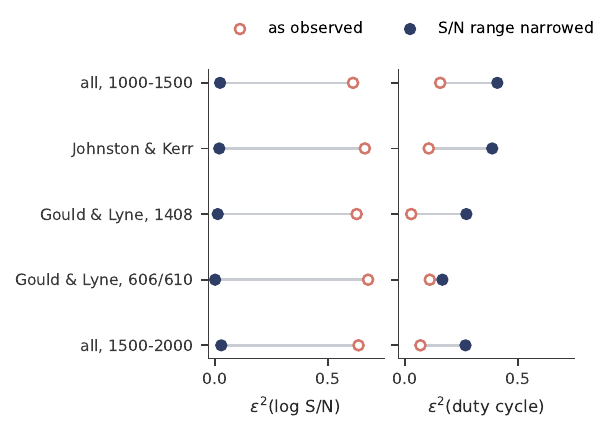}
\caption{EPN profiles: the same subset of each sample as observed (open) and after degrading every profile to the subset's lowest \snr (filled; Ward, $k = 5$, median of five noise realisations).
The partition stops following \snr (left, re-measured after degrading) and follows duty cycle (right).}
\label{fig:epn}
\end{figure}

To identify what drives the partition once \snr no longer sorts it, we bring the profiles of each sample to nearly the same \snr.
We proceed by adding white noise to every profile above a sample's $25$th-percentile \snr until it falls to that value.
We then drop the faintest quarter, since it cannot be raised to this level. 
Finally, we compare with the same profiles as observed (Fig.~\ref{fig:epn}). 
The samples then hold $1019$ profiles (all papers, 1000--1500\,MHz), $445$ \citep{jk18}, $206$ \citep[][1408\,MHz]{gl98} and $207$ \citep[][606/610\,MHz]{gl98}, with five noise realisations each.

The partition stops following \snr ($\eps \leq 0.03$ against the re-measured \snr and $\leq 0.01$ against the original) and catalogue flux density ($\eps \leq 0.05$), and follows duty cycle more strongly ($\eps$ rises by $0.06$ to $0.28$, to $0.16$--$0.41$).
The association does not come from pulses wider than the band, which account for $2$ to $7$ per cent of the profiles, but from DTW reducing differences in width without removing them, which the sort by \snr hides until the range is narrowed (Appendix~\ref{app:supp_case}).
On the real profiles, the association with duty cycle disappears once profiles are stretched to a common width ($\eps = 0.02$).

The partition follows width as a fraction of the rotation, and its association with $P$ ($\eps(\log P) = 0.26$) comes through the period--width relation \citep{rankin1990}, since it disappears once the duty cycle is held fixed ($\eps = 0.02$).
The partition is also associated with $\dot{E}$ in the all-papers sample at $1000$--$1500$\,MHz ($\eps(\log\dot{E}) = 0.11$, against $0.03$ before narrowing), which brightness cannot explain ($\rho^{2}\,\eps(\log\snr) < 0.001$; Table~\ref{tab:case_spin}).
This association also comes through $P$, since it disappears once $P$ is held fixed ($\eps \leq 0.01$ in all four samples).

\begin{table*}
\caption{Spin-down quantities after narrowing the \snr range (DTW, Ward, $k = 5$, medians of five noise realisations).
$\eps(\log\dot{E})$ before is for the same profiles as observed.
$\tau_{\mathrm{c}} = P/(2\dot{P})$ is the characteristic age, and $\dot{E} \mid P$ is the residual of a straight-line fit of $\log\dot{E}$ on $\log P$.
$\rho(\log\dot{E}, \log\snr)$ uses the \snr as observed, and $\rho^{2}\,\eps(\log\snr)$ is the part of $\eps(\log\dot{E})$ expected from brightness alone (Sect.~\ref{sec:null}).}
\label{tab:case_spin}
\setlength{\tabcolsep}{3pt}
\begin{tabular}{lrrrrrrrr}
\toprule
 & & \multicolumn{2}{c}{$\eps(\log\dot{E})$} & & & & & \\
sample & $n$ & before & after & $\eps(\log P)$ & $\eps(\log\tau_{\mathrm{c}})$ & $\eps(\dot{E} \mid P)$ & $\rho(\log\dot{E}, \log\snr)$ & $\rho^{2}\,\eps(\log\snr)$ \\
\midrule
all papers, 1000--1500\,MHz  & 1019 & 0.03 & 0.11 & 0.26 & 0.06 & 0.00 & $0.12$ & $<0.001$ \\
\citet{jk18}, 1369\,MHz      & 445  & 0.01 & 0.05 & 0.10 & 0.02 & 0.00 & $0.04$ & $<0.001$ \\
\citet{gl98}, 1408\,MHz      & 206  & 0.06 & 0.03 & 0.07 & 0.02 & 0.01 & $0.19$ & $<0.001$ \\
\citet{gl98}, 606/610\,MHz   & 207  & 0.03 & 0.04 & 0.07 & 0.00 & 0.00 & $0.24$ & $0.001$ \\
\bottomrule
\end{tabular}
\end{table*}
Structure in profile width was therefore present throughout, hidden by the differences in sensitivity.
The mean distance still follows the re-measured \snr ($\rho \approx -0.40$ to $-0.48$), but follows the original \snr only at $\rho = -0.20$ and the catalogue flux density at about $-0.05$.
Controls with shuffled noise levels and with the bright profiles alone show that the narrower range, not the added noise, reveals this structure (Appendix~\ref{app:supp_case}).
The exception is \citet{gl98}, whose bright profiles still sort by \snr because they are so alike that a fresh noise realisation moves a profile by as much as the differences between profiles, so for DTW no fixed \snr threshold makes a sample safe.

\subsection{The selection and the models of ToPP~I}
\label{sec:case_toppI}

\citetalias{toppI} was primarily a methods paper, built on archival profiles whose data reduction it noted to be heterogeneous. 
Physical trends among the data and their relation to MSTs were therefore explored under this caveat. 
In this section, we test whether its selection and procedure are exposed to the sensitivity sort.

\citetalias{toppI} kept pulsars with a profile in all four bands and $\snr > 20$ in each, and excluded those with an obvious scattering tail at the lowest frequency, a visual judgement that we do not repeat.
The $116$ such pulsars in the EPN database (\citetalias{toppI} had $90$) still span a factor of $15$ to $80$ in \snr per band.
Of the $116$, $85$ are among the $90$ of \citetalias{toppI}, and the other $31$ are fainter and more dispersed (median dispersion measure $123$ against $50$\,pc\,cm$^{-3}$), but the results below hold on the $85$ alone (Appendix~\ref{app:supp_case}).

Applying DTW and Ward linkage to each band, the mean distance follows \snr in all four ($\rho = -0.60$ to $-0.74$), and with five groups the partition follows \snr at $\eps = 0.54$--$0.70$ in three of them.
At $400$--$700$\,MHz it puts $108$ of the $116$ pulsars in one group ($\eps = 0.05$), but with eight groups it follows \snr at $\eps = 0.56$.
In three bands it carries no association with $\dot{E}$ ($\eps \leq 0.01$).
In the fourth, $1500$--$2000$\,MHz, where $\dot{E}$ follows \snr ($\rho = 0.41$), the partition is associated with $\log\dot{E}$ at $\eps = 0.17$ ($p = 4\times10^{-5}$), which falls to $0.003$ once a least-squares straight line in $\log\snr$ is subtracted from $\log\dot{E}$, and to $0.000$ when the mean of $\log\dot{E}$ in each tenth of \snr is subtracted instead.
Taken on its own, this association would be reported as physical, but almost all of it is the part of $\dot{E}$ that follows \snr.

Splitting the band by brightness cannot settle whether any of the remainder is real given the association is $0.02$ in the fainter half and the brighter half falls almost entirely into a single group ($55$ of $58$ profiles).
Moreover, $111$ of the $116$ profiles in this band come from \citet{gl98}, the exceptional sample of Sect.~\ref{sec:case_narrow}.

In practice, \citetalias{toppI} compared Gaussian-mixture models of the profiles rather than the noisy profiles themselves, and interpreted the DTW distance between a pulsar and the root of the MST, $w_{ri}$, as the complexity of its profile.
Under this DTW, pairs of fainter observed profiles are further apart than pairs of brighter ones (median distance $33$ against $18$), so the noise pushes faint profiles away from one another, as in Sect.~\ref{sec:mechanism}, rather than making them alike.
On the published tree of \citetalias{toppI}, fainter pulsars are no more often leaves than brighter ones ($19$ of $45$ against $21$ of $45$; Fisher's exact test $p = 0.83$), and the complexity does not follow \snr ($\rho = -0.02$).
Any sensitivity sort in the published tree is therefore at most about $|\rho| = 0.2$. The $95$\% intervals are $-0.23$--$0.19$ for $\rho$, and $-0.25$--$0.16$ for the difference in leaf fractions between fainter and brighter pulsars.

We repeated the procedure of \citetalias{toppI} with its original code and in its order, from re-gridding to $512$ bins and scaling to $[0, 1]$, through aligning the bands and fitting the main pulse and interpulse separately, to DTW and the MST, on the same $90$ pulsars and references.
The mixture fit has no fixed random seed, so its models cannot be reproduced exactly. 
The re-run shares $29$ of the $89$ edges with the published tree.
The order of $w_{ri}$ is nevertheless recovered at rank correlation $0.96$ with the published order.
Once again, fainter and brighter pulsars are leaves just as often ($17$ of $45$ against $19$ of $45$; $p = 0.83$), and the complexity follows \snr at most weakly ($\rho = -0.21$, $p = 0.05$).
The result of \citetalias{toppI} is thus stable when the models are refitted, although single edges of the tree are not.

This independence from \snr depends on the model.
Fitting each profile with a least-squares sum of at most eight Gaussians and choosing their number with the Bayesian information criterion, we obtain a median of four components.
Both the tree and the complexity follow \snr: $31$ of the $45$ fainter pulsars are leaves against $17$ of the $45$ brighter ones ($p = 0.006$), and the complexity follows \snr at $\rho = -0.34$ ($p = 0.001$).
Whether the procedure carries \snr into its result therefore depends on the models (see Appendix~\ref{app:mst_epn} for a comparison of models and distances on the \citetalias{toppI} sample).

\citetalias[Section~3.2]{toppI} reported a Pearson correlation of $0.34$ between complexity and characteristic age, $P/(2\dot{P})$.
On the published tree the rank correlation is $0.28$ ($p = 0.008$), and since age is not tied to \snr in this sample ($\rho = -0.04$), brightness does not produce it.
It comes instead from the six millisecond pulsars, which are old, mostly bright in this sample (\snr percentiles $18$ to $91$, median $72$) and have distinctive profiles, and without them it falls to $0.14$ ($p = 0.22$); the re-run of the original procedure gives $0.27$ and $0.12$.
Complexity is not tied to $\dot{E}$ either ($\rho = -0.10$, $p = 0.36$).
What the complexity separates is millisecond pulsars from the rest, not young from old.

\section{Discussion and recommendations}
\label{sec:discussion}

Nothing here is specific to pulsars.
A sample is affected when two conditions are met. 
Firstly, the dissimilarity between objects must correspond to the case where noise adds more distance than the real differences between them. 
Secondly, the sensitivity must vary across the sample and correlate with a given property of interest---which occurs in nearly every flux-limited survey.
The property at risk is whichever one tied to brightness: $\dot{E}$ for pulsars, absolute magnitude or stellar mass for galaxies, peak luminosity for transients, and distance everywhere.

DTW meets the first requirement because it makes the distance less sensitive to differences in width and component spacing but not to noise (Sect.~\ref{sec:metrics}). 
Under hierarchical clustering, the pointwise distances we tested generally do not, including correlation distance, which rescales each profile. 
Spectral clustering, however, finds the artefact with pointwise distances too, in five of twelve variants (Appendix~\ref{app:supp_metrics}).

The direction of the artefact is predictable, since faint objects look more complex by their distance to the root, though not in the number of fitted components, and sit at the edge of the graph, and it usually has a plausible physical story attached, which in a sense is what makes it dangerous.
Sensitivity can also act without noise, through an instrument that shapes the measured features directly, such as a beam whose shape changes across the band \citep{vohl2026-frb}.

Orderings read from an MST are also exposed.
Run unchanged on the null population of Sect.~\ref{sec:null}, the Sequencer of \citet{baron2021} recovers the number of components, but within each class of components it orders objects by \snr, at a rank correlation between $-0.30$ and $-0.51$ in each of ten realisations, against $-0.14$ to $0.10$ when every object has the same \snr.
A single Euclidean MST walked in the same way does the same, so the cause is the tree, not the choice of distances and scales.
The trends that \citet{baron2021} report in the Sloan spectra, with stellar temperature and with redshift, are both tied to brightness in a flux-limited sample, although redshift moves spectral features much further than noise inflation is likely to.

The published tree and complexity of \citetalias{toppI} are free of the sensitivity sort, but its age relation comes from the millisecond pulsars, and the association with $\dot{E}$ in the public profiles comes from brightness or, once the \snr range is narrowed, from the spin period through the period--width relation.
Results in this paper are methodological, testing previous claims for pulsar physics.
Based on these results, we suggest the following steps to an author about to classify a sample of unequal sensitivity:
\begin{enumerate}
\item Measure the artefact directly on a simulation matched to the pipeline, as the partition's effect size against \snr when the property follows only \snr, centring the simulated profiles as the real ones are centred, and if the distance warps, vary the band.
$\mathcal{R}$ (Eq.~(\ref{eq:ratio})) describes the mechanism but does not rank moderate configurations.
To compare shape without width, rescale each profile to a common width and compare pointwise rather than warp; a pointwise distance on unscaled profiles compares width as well.
Treat spectral clustering with caution.
\item Report $\eps(\log\snr)$ alongside $\eps(y)$, and compare $\eps(y)$ with $\rho^{2},\eps(\log\snr)$, the part expected from brightness alone.
A relation that survives shuffling $y$ among objects of similar \snr is more credible, though a selection on \snr can still pass.
Ensure that $\rho$ on the sample is measured, as it varies widely between them (Sect.~\ref{sec:case_toppI}).
\item Report the relation in several brightness bins, idealy with uncertainties, interpreting a relation as real only if it strengthens toward the bright end by more than its sampling error.
\item Inject ground truth across a range of noise, not at a single level.
\item Test the partition against known instrumental variables, not only \snr.
\end{enumerate}

\section{Conclusions}
\label{sec:conclusions}

\begin{enumerate}
\item Distance-based unsupervised classification of observations of unequal sensitivity can partition by sensitivity and then report significant associations between morphology and a quantity it does not depend on; the effect size scales as $\rho^{2}\,\eps(\log\snr)$, and in our simulations a coupling of $\rho = 0.26$ gives $\eps \simeq 0.06$, still significant in half of the realisations. 
\item Warping is at cause, as it makes the distance less sensitive to differences in width, component spacing, and to amplitude ratio of two components, which does not depend on width, but not to noise; pointwise distances, also on profiles rescaled to a common width, do not turn the inflation into groups under hierarchical clustering; a minimum spanning tree built from them can still place faint objects at its leaves (Sect.~\ref{sec:discussion}).
\item With wide-band warping, checking whether the groups follow \snr flags every compromised analysis but cannot reject a relation;
splitting by brightness gives false negatives on a real relation strongly tied to brightness, and false positives on a weak artefact unless a margin is required; smoothing and derivative warping remove the artefact only with the morphology; and bootstrap, fresh-noise and permutation controls pass it.
\item In public pulse profiles every sample is sorted by \snr under wide warping and none under a pointwise comparison at a common width, which also removes the duty cycle; width (duty-cycle) structure appears once the \snr range is narrowed.
In an earlier classification, the published tree and complexity do not follow \snr and a re-run of its pipeline recovers the complexity ranking, though not single edges of the tree, but other models can carry \snr into the tree.
\end{enumerate}

No new physical relation follows from these results beyond the known period--width relation; relations with age or $\dot{E}$ found by such classifications should be claimed only from analyses that pass the checks of Sect.~\ref{sec:discussion}.

\section*{Data availability}
The simulations use no observational data; Sect.~\ref{sec:case} uses the EPN database and the ATNF catalogue \citep{atnf}.
A reproduction package will be made available on Zenodo on publication.
The random seeds behind each simulated result are listed in Appendix~\ref{app:seeds}.

\begin{acknowledgements}
We thank Michael Keith for maintaining access to the EPN database.
This research has made use of the ATNF Pulsar Catalogue \citep{atnf}.
Software: Astropy \citep{astropy:2013, astropy:2018, astropy:2022}, dtw-python \citep{giorgino2009computing}, joblib, Matplotlib \citep{Hunter:2007}, NetworkX \citep{hagberg2008}, Numba \citep{lam2015numba}, NumPy \citep{harris2020array}, pandas \citep{reback2020pandas}, psrqpy \citep{psrqpy}, scikit-learn \citep{scikit-learn}, SciPy \citep{virtanen2020scipy} and the Sequencer \citep{baron2021}.
Claude Code (Anthropic; models: Haiku 5.5, Opus 5, and Opus 5.5) was used between August and November 2026 under DV's direction and oversight to write and run analysis code, including the simulations and the re-run of the ToPP I procedure; to produce figures; to revise the manuscript, check its references, and check its numbers against the code outputs.
\end{acknowledgements}

\bibliographystyle{format/aa}
\bibliography{refs}

\begin{appendix}
\input{appendix}

\end{appendix}

\end{document}

%% file: appendix.tex
\section{Minimum spanning trees of the profiles classified by ToPP~I}
\label{app:mst_epn}

To show what the procedure of \citetalias{toppI} does to individual pulsars, we build the MST of its $90$ pulsars under that procedure and under four pipelines that each change one step of it.
All five use the same profiles, prepared and aligned as in \citetalias{toppI} (Sect.~\ref{sec:case_toppI}), and sum the distances over the four frequency bins.

The procedure of \citetalias{toppI} compares Gaussian-mixture models of the profiles with DTW, using an asymmetric step pattern and a band of $256$ of the $512$ samples, taking the larger of the two asymmetric distances and the smaller of the direct and phase-reversed comparisons.
The other four pipelines apply this distance to the observed profiles and to the least-squares models of Sect.~\ref{sec:case_toppI}, and Euclidean distance to both.
Since the mixture fit removes the baseline by clipping each profile at its median plus three times its noise, the observed profiles have their median subtracted, and the least-squares models are fitted to these.

No morphological truth is known for these pulsars, so we score each tree on what it should and should not follow.
For shape we use the duty cycle, $W_{50}/P$, from the ATNF catalogue (known for $89$ of the $90$ pulsars, and only weakly tied to the mean $\log\snr$, $\rho = -0.19$).
The edge agreement in a quantity is the Spearman rank correlation between its values at the two ends of each edge, tested against $2000$ permutations.
A tree organised by shape should show edge agreement in duty cycle but not in \snr, no correlation $\rho_{d}$ between a pulsar's mean distance to all others and its \snr (Sect.~\ref{sec:mechanism}), and similar fractions of leaves among fainter and brighter pulsars.

\begin{figure*}
\centering
\includegraphics[width=\textwidth]{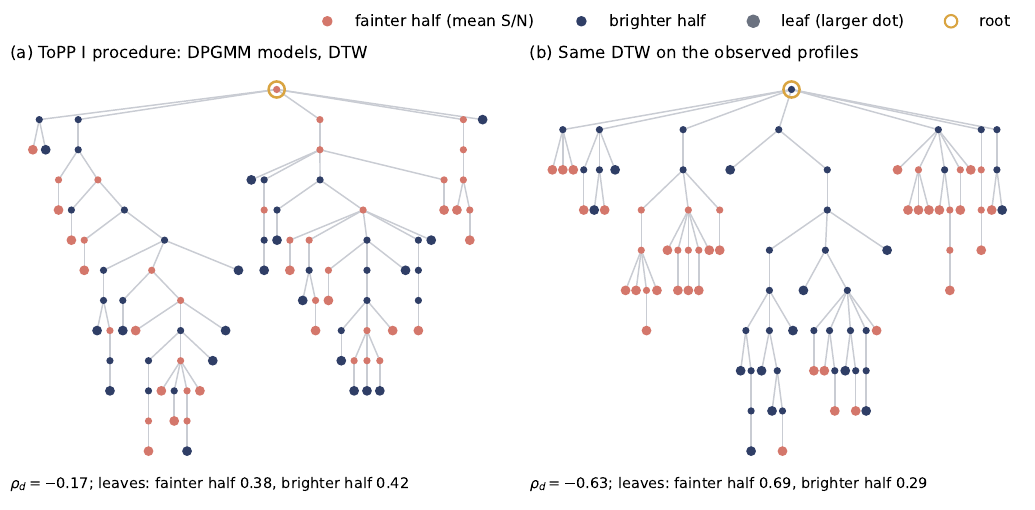}
\caption{Minimum spanning tree of the $90$ pulsars of \citetalias{toppI} (a) under the procedure of \citetalias{toppI}, Gaussian-mixture models compared by its DTW, and (b) under the same DTW applied to the observed profiles.
Each tree hangs from its root, the pulsar of highest closeness centrality, and every other pulsar sits at a depth given by its number of edges from the root.
Pulsars are coloured by whether their mean $\log\snr$ over the four bins lies below or above the median, and leaves are drawn larger.
In (b), the fainter pulsars collect as leaves.}
\label{fig:mst_epn_toppI}
\end{figure*}

The procedure of \citetalias{toppI} gives the tree least tied to sensitivity (Fig.~\ref{fig:mst_epn_toppI} and Table~\ref{tab:mst_epn}).
It passes all four criteria, and its root is J0614+2229, the root of the published tree.
The Gaussian-mixture and least-squares fits both keep a median of four components per profile.
The same distance on the observed profiles fails all three criteria tied to \snr, with faint pulsars at the edges (Fig.~\ref{fig:mst_epn_toppI}b). 
Thus, in this sample the Gaussian-mixture models remove the sensitivity that the distance would otherwise carry into the tree.
Clipping the baseline alone, as the mixture fit does beforehand, removes most of the sort from the mean distance and the edges, although fainter pulsars remain leaves slightly more often ($0.56$ against $0.33$, $p = 0.06$).
With least-squares models, faint pulsars remain at the edges.
Euclidean distance shows no edge agreement in \snr, but the mean distance follows \snr weakly in both cases; on observed profiles, faint pulsars sit at the edges.
Duty cycle is a property of the shape, not the shape itself. 
Agreement in duty cycle shows that a tree follows some physical property, not that it orders profiles by their full shape.

\begin{table}[h!]
\centering
\caption{Minimum spanning trees of the $90$ pulsars of \citetalias{toppI} under five pipelines.
$\rho_{d}$ is the Spearman rank correlation between a pulsar's mean distance to all others and its mean $\log\snr$ over the four bins.
Leaves give the fraction of leaves in the fainter and the brighter half.
The last two columns give the edge agreement in $\log(W_{50}/P)$ and in $\log\snr$; all have $p \leq 0.0005$ except those marked $^{a}$ ($p = 0.12$), $^{c}$ ($p = 0.08$) and $^{b}$ ($p > 0.7$).}
\label{tab:mst_epn}
\footnotesize
\setlength{\tabcolsep}{3pt}
\begin{tabular}{@{}>{\raggedright\arraybackslash}p{3.3cm}rcrr@{}}
\toprule
pipeline & $\rho_{d}$ & leaves & $W_{50}/P$ & $\snr$ \\
\midrule
Gaussian mixture, DTW of \citetalias{toppI}  & $-0.17$ & 0.38, 0.42 & $0.68$ & $0.16^{a}$ \\
observed, DTW of \citetalias{toppI} & $-0.63$ & 0.69, 0.29 & $0.45$ & $0.68$ \\
least squares, DTW of \citetalias{toppI} & $-0.26$ & 0.69, 0.38 & $0.59$ & $0.17^{a}$ \\
observed, Euclidean           & $-0.29$ & 0.64, 0.38 & $0.66$ & $-0.01^{b}$ \\
least squares, Euclidean      & $-0.24$ & 0.49, 0.40 & $0.70$ & $0.03^{b}$ \\
observed, clipped as for the mixture, DTW of \citetalias{toppI} & $-0.17$ & 0.56, 0.33 & $0.71$ & $0.19^{c}$ \\
\bottomrule
\end{tabular}
\end{table}

\section{Supporting tests}
\label{app:supp}

This appendix gives the details behind results that the body states in one sentence, in the order of the body.

\subsection{Realisations}
\label{app:seeds}

Every simulated result depends on random draws: the \snr, components, properties and noise of each object, and the starting points of $k$-medoids and spectral clustering.
A realisation is one set of draws, fixed by the seed of the random number generator in the released code.
Wherever the body shows one realisation, it is seed $7$; ranges over ten realisations use seeds $7$ to $16$, and the Gaussian-mixture fits, which use a Dirichlet-process Gaussian mixture, use seeds $7$ to $11$.
Seed $7$ gives the second-smallest Ward artefact of the ten, but the largest under spectral clustering.

\subsection{The failure mode and the null population}
\label{app:supp_null}

With a band of zero width the DTW path is fixed, and the expected excess per sample is the mean of a folded normal, which matches the measured inflation of Sect.~\ref{sec:mechanism} at a rank correlation of $0.99$.
The intermediate \snr ranges $20$--$200$, $40$--$120$ and $60$--$90$ give median correlations of distance with \snr of $-0.79$, $-0.60$ and $-0.40$.
The sample-size test uses ten realisations per size at $256$ samples.
Variants with component widths $w = 3$ and $12$ samples are among the twelve of Appendix~\ref{app:supp_metrics}.

With the object's own noise-free profile as the model, the floor includes its real distance to the others; the correction removes the signal with the noise, taking the correlation of the mean distance with \snr from $-0.87$ to $-0.02$ and with the true number of components from $0.22$ to $0.00$ (twenty null realisations).
The noise part of the floor alone, the difference between the floors of a noisy and a noise-free single-Gaussian fit, keeps the component signal ($0.35$, against $0.47$ without noise, for profiles of random width) with no correlation with \snr ($-0.03$).

\subsection{Which distances are vulnerable}
\label{app:supp_metrics}

Twelve variants of the population each change one assumption, or in two cases the warping band, with five realisations each.
DTW has both the largest $\mathcal{R}$ and the largest artefact in all twelve, but the rank correlation between $\mathcal{R}$ and $\eps(y)$ is only $0.55$, and $0.34$ within DTW.
In one variant the coupling is weaker, $\rho \simeq 0.3$, and DTW gives $\eps(y) = 0.075$.
Spectral clustering produces an artefact with all three pointwise distances in five of the twelve.

Centring each profile on its noisy peak after a random rotation creates no artefact for the pointwise distances ($\eps \leq 0.01$; medians of ten realisations).
Euclidean distance then keeps only $0.04$ of its $0.22$ for the genuine relation and $0.25$ of its $0.55$ for the asymmetry relation, the stretched comparison keeps $0.19$ and $0.41$, and DTW, which is insensitive to the shift, $0.03$ and $0.06$.
The stretched comparison measures the width from the peak to the first half-maximum crossing on each side.
Using the outermost crossings, as in the catalogue $W_{50}$, keeps the asymmetry relation ($\eps = 0.79$) with no artefact ($\eps(\log\snr) \leq 0.01$).

In Table~\ref{tab:remedies}, smoothing uses a Gaussian kernel whose standard deviation is given in samples.
Clipping is as in the Gaussian-mixture fits of Appendix~\ref{app:mst_epn}, and the on-pulse window keeps the samples where a copy smoothed over two samples exceeds the clipping level.
Smoothing over ten samples cuts the genuine relation from $0.22$ to $0.09$ under Euclidean distance and from $0.28$ to $0.08$ under DTW without a band, as the components are only $w = 6$ samples wide (about $14$ at half maximum).
Coding the genuine relation in the number of components does not separate it from the artefact, because more components make a wider profile.
The partition then follows the width ($\eps = 0.79$), and with the width randomised no distance recovers the number of components ($\eps \leq 0.19$).

\begin{table}
\centering
\caption{Smoothing, derivative warping and suppression of the off-pulse region applied to DTW (band half-width $38$) on the two-property population of Sect.~\ref{sec:populations} (median of seeds $7$--$11$; Table~\ref{tab:band} gives the first row over ten realisations).}
\label{tab:remedies}
\begin{tabular}{lrr}
\toprule
variant & $\eps$(genuine) & $\eps$(spurious) \\
\midrule
none                               & 0.03 & 0.45 \\
smoothing over 2 samples           & 0.13 & 0.42 \\
smoothing over 5 samples           & 0.08 & 0.30 \\
smoothing over 10 samples          & 0.03 & 0.11 \\
derivative DTW                     & 0.00 & 0.42 \\
derivative, smoothed over 2        & 0.19 & 0.38 \\
derivative, smoothed over 5        & 0.10 & 0.30 \\
derivative, smoothed over 10       & 0.13 & 0.26 \\
baseline clipped as in \citetalias{toppI} & 0.16 & 0.24 \\
on-pulse window only       & 0.18 & 0.14 \\
\bottomrule
\end{tabular}
\end{table}

\begin{figure*}[h!]
\centering
\includegraphics[width=0.85\textwidth]{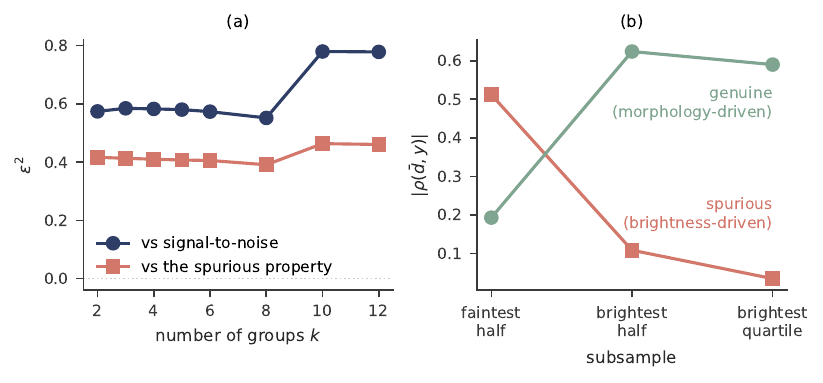}
\caption{Two checks on the two-property population (DTW, Ward).
(a) Effect size of the partition against $\log\snr$ and against the property tied to brightness, for 2 to 12 groups $k$ (least favourable of ten realisations).
(b) $|\rho(\bar{d}, y)|$, the absolute rank correlation of the mean distance with the property tied to brightness and with the genuine property, in the faintest half, the brightest half and the brightest quarter (one realisation).}
\label{fig:diagnostic}
\end{figure*}

\subsection{Diagnostics}
\label{app:supp_diag}

With $120$ objects at $128$ samples, the DTW partition keeps more of a genuine relation not tied to brightness (median $\eps(y) = 0.16$, against $0.01$ at $180$ objects and $256$ samples).

With a band of $4$ or $8$ samples only, the check of Sect.~\ref{sec:power} flags $77$ to $100$ of $100$ realisations irrespective of the ground truth. 
With a pointwise distance and brightness alone, it mostly compares two effect sizes below $0.01$.

Table~\ref{tab:strat_mech} also tests two further kinds of relation.
In the selection case, the property and the component separation are independent, but \snr rises with the property and falls with the separation, since at fixed flux density a wider profile has a lower peak.
In a relation confined to faint objects, the property sets the component separation only below the median \snr.
Signs change across brightness bins for artefacts and real relations alike.
The rates are over $300$ realisations, since the first $100$ and the next $200$ agree within binomial chance.
Comparing halves or quarters of equal size instead gives $34$ and $53$ per cent false positives on the weaker artefact, against $45$ per cent, and requiring a bootstrap interval to exclude zero misses $57$ to $100$ per cent of real relations.

The shuffled tests of Table~\ref{tab:strat_mech} use $999$ shuffles within twentieths of \snr, and for the mean distance the observed correlation is measured from the centre of the shuffled ones, since the shuffles keep the part of the correlation that comes through \snr.

\begin{table*}
\centering
\caption{Splitting by brightness on five kinds of relation (Sect.~\ref{sec:strat}; seeds $7$--$306$).
Relation: type of relation; truth: artefact or real; $\rho_{y}$: median $\rho(y, \log\snr)$; wrong: error rate of the check over $300$ realisations; margin: the same with a margin of $0.15$; faint, bright: median $\rho(\bar{d}, y)$ in the faintest half and the brightest quarter.
Plain, shuf.\ $\bar{d}$, shuf.\ part.: errors in the first $100$ realisations at $p < 0.01$ for the Kruskal--Wallis test of the partition, and with $y$ shuffled within twentieths of \snr, tested against $\bar{d}$ and against the partition.
Rates are in per cent with $95$ per cent binomial intervals, and $\rho_{y}$ uses the first $100$ realisations.}
\label{tab:strat_mech}
\footnotesize
\setlength{\tabcolsep}{2.5pt}
\begin{tabular}{lllrrrrrr}
\toprule
relation & truth & $\rho_{y}$ & wrong & margin & faint, bright & plain & shuf.\ $\bar{d}$ & shuf.\ part. \\
\midrule
brightness only    & artefact & 0.71 & 1 (0--2) & 0 (0--1) & $-0.43$, $-0.02$ & 100 & 0 & 2 \\
brightness only    & artefact & 0.30 & 45 (39--50) & 7 (4--10) & $-0.15$, $0.01$ & 59 & 0 & 2 \\
real, not tied     & real     & $-0.01$ & 11 (8--15) & 33 (28--39) & $0.10$, $0.34$ & 76 & 5 & 73 \\
real and tied      & real     & 0.72 & 74 (69--79) & 93 (89--95) & $-0.40$, $0.29$ & 0 & 11 & 80 \\
real and tied      & real     & 0.30 & 10 (7--14) & 34 (29--40) & $-0.07$, $0.33$ & 25 & 5 & 78 \\
selection           & artefact & 0.60 & 6 (4--10) & 0 (0--1) & $-0.33$, $0.07$ & 100 & 79 & 12 \\
real, faint only   & real     & $-0.01$ & 43 (38--49) & 85 (81--89) & $0.08$, $-0.03$ & 85 & 98 & 82 \\
\bottomrule
\end{tabular}
\end{table*}

\subsection{Public pulse profiles}
\label{app:supp_case}

Over three to eight groups, the partition's effect size against \snr ranges from $0.23$ to $0.86$, the lowest where a \citet{gl98} sample splits into one large group and two small ones.
The correlation of the mean distance with $\log\dot{E}$ is stronger in the brightest quarter than in the faintest half in four of the seven samples; this would be interpreted as real by the check of Sect.~\ref{sec:strat}. 
However, with such weak couplings, it reads a pure artefact as real in about two of five cases.
Only \citet{gl98} at 606/610\,MHz exceeds the margin of $0.15$ ($-0.01$ to $-0.34$), and its significance ($p = 0.02$) does not survive seven tests.
In the bright \citet{gl98} profiles, a fresh noise realisation moves each profile by $1.0$ times the interquartile range of the distances between profiles, against $0.39$ to $0.91$ in the other samples.

At five groups, native bin count gives $\eps = 0.02$--$0.16$, dispersion measure $0.00$--$0.12$ and catalogue scattering time $0.01$--$0.19$ across the seven samples, against $0.26$--$0.85$ for \snr.
Observing frequency gives $\eps = 0.15$ at $700$--$1000$\,MHz, through \snr, and $0.26$ in the \citet{gl98} sample at $606/610$\,MHz, where no pulsar has profiles at both frequencies, so the label separates two sets of observations with median \snr of $25$ and $103$, and about half of the association comes through \snr; elsewhere it is at most $0.04$.
The group by publication table gives Cram\'er's $V = 0.17$--$0.36$, against $0.12$--$0.21$ with the publications shuffled.
\snr is only weakly tied to duty cycle ($\rho = -0.29$ to $0.01$), which at fixed flux density sets the peak \snr \citep{lorimer2004}, and to dispersion measure ($-0.34$ to $0.00$).
Profiles from \citet{gl98} do carry correlated off-pulse noise (lag-one autocorrelation $0.53$--$0.55$ at $256$ bins, against $0.02$ for \citet{jk18}), probably from smoothing before publication---corresponding to a case covered by the correlated-noise variant of Sect.~\ref{sec:metrics}.
Equalising with added noise that has this autocorrelation instead of white noise leaves the results unchanged over $40$ noise realisations, and the exception of the bright \citet{gl98} profiles does not depend on the noise type either.

In synthetic populations with random widths and every object at one \snr ($25$ or $100$), the DTW partition (band half-width $38$) follows width at $\eps = 0.21$--$0.29$, against $0.43$--$0.62$ for Euclidean distance (medians of ten realisations).
On the $85$ pulsars shared with \citetalias{toppI} (Sect.~\ref{sec:case_toppI}), the mean distance follows \snr in all bands ($\rho = -0.51$ to $-0.70$), and at $1500$--$2000$,MHz the association with $\log\dot{E}$ falls from $0.15$ to $0.007$ once a linear fit in $\log\snr$ is removed.

At phase sampled over $512$ bins, the $1359$ profiles of the $1000$--$1500$\,MHz sample are sorted by \snr as strongly as with $256$ phase bins, although $453$ of them are interpolated.
The mean distance follows \snr at $\rho = -0.78$, and the Ward partition gives $\eps(\log\snr) = 0.52$, $\eps(\log\dot{E}) = 0.03$ and $\eps(\mathrm{duty}) = 0.19$, against $0.54$, $0.02$ and $0.09$ at $256$ samples.

With the outermost half-maximum crossings, as in the catalogue $W_{50}$, the stretched comparison keeps more of the duty cycle ($\eps = 0.08$--$0.21$) but follows \snr more ($0.01$--$0.16$), since at low \snr a noise spike can add a false outer crossing; it still sorts less than DTW in every sample.
The comparison at a common width drops $34$ pulsars without a usable catalogue width ($32$ have no \texttt{W50} entry in version 2.8.1, and two are listed as $0.0$\,ms), $33$ of them millisecond pulsars, whose \snr does not differ from the rest; in the largest sample, dropping them lowers $\eps(\log\snr)$ for DTW from $0.54$ to $0.41$.

Adding the noise levels in shuffled order gives the same result, so the narrower range is what matters.
Without any added noise, the bright profiles alone ($\snr \geq 100$ at $1000$--$1500$\,MHz) display the same duty-cycle structure ($\eps = 0.34$--$0.38$).

\begin{table}
\centering
\caption{EPN profiles rescaled to a common width ($n$ excludes the pulsars without a catalogue width; Ward, $k = 5$).
Columns give $\eps(\log\snr)$, except where marked: DTW and Euclidean distance on the profiles as observed, and on profiles stretched to a common half-maximum width.
JK18 is \citet{jk18}, GL98 is \citet{gl98}, and all combines every publication in the band; $^{a}$~$\eps(\mathrm{duty})$ of the Euclidean partition.}
\label{tab:case_width}
\footnotesize
\setlength{\tabcolsep}{3pt}
\begin{tabular}{llrrrrrr}
\toprule
 & & & \multicolumn{3}{c}{as observed} & \multicolumn{2}{c}{common width} \\
sample & MHz & $n$ & DTW & Eucl. & duty$^{a}$ & DTW & Eucl. \\
\midrule
all                & 1000--1500 & 1325 & 0.41 & 0.05 & 0.54 & 0.51 & 0.01 \\
JK18               & 1369       & 594  & 0.64 & 0.11 & 0.61 & 0.37 & 0.01 \\
GL98               & 1408       & 275  & 0.81 & 0.02 & 0.46 & 0.30 & 0.00 \\
GL98               & 606/610    & 276  & 0.57 & 0.08 & 0.35 & 0.23 & 0.03 \\
all                & 1500--2000 & 326  & 0.67 & 0.00 & 0.39 & 0.22 & 0.04 \\
\bottomrule
\end{tabular}
\end{table}

\subsection{Sequencer preprocessing and open cases}
\label{app:supp_disc}

The elongation of the Sequencer's tree \citep{baron2021} falls as the noise rises, so a short tree flags noisy data, but a long one does not show that the order is physical.
The Sequencer package accepts only positive inputs for the Kullback--Leibler divergence, and non-negative inputs for the Earth mover and energy distances. 
Our noisy peak-normalised profiles have negative bins and need preprocessing.
We obtained the numbers of Sect.~\ref{sec:discussion} by adding a constant independent of \snr to every object.
Clipping the negative bins leaves instead a positive off-pulse baseline offset of about $\sigma/\sqrt{2\pi}$ per bin that grows with the noise. 
Thus, profiles already differ by \snr before any distance is computed.
Finally, we note that we did not test the Earth mover distance on its own (although it tolerates shifts as warping does) nor non-linear embeddings such as UMAP and $t$-SNE, which the spectral-clustering result suggests deserve scrutiny.
In pulsars, scattering makes profiles simpler, which is the opposite of the artefact, and the mechanism may also apply to one object followed across epochs \citep{rajwade2022}.